\documentstyle[psfig,amssymb,pagenum]{llncs}

\begin{document}

\newcommand{\beq}{\begin{equation}}
\newcommand{\eeq}{\end{equation}}
\newcommand{\bea}{\begin{eqnarray*}}
\newcommand{\eea}{\end{eqnarray*}}
\newcommand{\eps}{\epsilon}
\newcommand{\mat}{\left( \! \begin{array}{rr}}
\newcommand{\rix}{\end{array} \! \right)}
\newcommand{\cho}{\left( \! \begin{array}{c}}
\newcommand{\ose}{\end{array} \! \right)}
\newcommand{\QED}{\rule{3mm}{3mm}}
\newcommand{\init}{{\mbox{\scriptsize{init}}}}
\newcommand{\accept}{{\mbox{\scriptsize{accept}}}}
\newcommand{\ket}{\rangle}
\newcommand{\bra}{\langle}
\newcommand{\back}{\backslash}
\newcommand{\R}{{\Bbb R}}
\newcommand{\C}{{\Bbb C}}
\newcommand{\To}{\Rightarrow}
\newcommand{\of}{\overline{f}}
\newcommand{\oq}{\overline{q}}
\newcommand{\oQ}{\overline{Q}}
\newcommand{\opi}{\overline{\pi}}
\newcommand{\oeta}{\overline{\eta}}
\newcommand{\oM}{\overline{M}}
\newcommand{\id}{{\bf 1}}

\title{Quantum Automata and Quantum Grammars}
\author{Cristopher Moore and James P. Crutchfield}
\institute{Santa Fe Institute, 1399 Hyde Park Road, Santa Fe NM 87501 USA \\
{\sf \{moore,jpc\}@santafe.edu}}
\maketitle

\begin{abstract}
To study quantum computation, it might be helpful to generalize
structures from language and automata theory to the quantum case.
To that end, we propose quantum versions of finite-state and push-down
automata, and regular and context-free grammars.  We find analogs of 
several classical theorems, including pumping lemmas, closure properties, 
rational and algebraic generating functions, and Greibach normal form.
We also show that there are quantum context-free languages that are not
context-free.
\end{abstract}

\section{Introduction}

Nontraditional models of computation --- such as real-valued, analog,
spatial, molecular, stochastic, and quantum computation --- have
received a great deal of interest in both physics and computer science
in recent years
(e.g.\ \cite{adleman,bss,deutsch,moore,Crut92c,siegel,Crutchfield&Mitchell94a}).
This stems partly from a desire to understand computation in dynamical
systems, such as ordinary differential equations, iterated maps,
cellular automata, and recurrent neural networks, and partly from a
desire to circumvent the fundamental limits on current computingx
technologies by inventing new computational model classes.

Quantum computation, in particular, has become a highly active research area.
This is driven by the recent discovery of quantum algorithms for factoring
that operate in polynomial time \cite{shor}, the suggestion that quantum
computers can be built using familiar physical systems
\cite{cirac,gershenfeld,lloyd}, and the hope that errors and decoherence
of the quantum state can be suppressed so that such computers can operate
for long times \cite{shor2,steane}.  

If we are to understand computation in a quantum context, it might be
useful to translate as many concepts as possible from classical
computation theory into the quantum case. From a practical viewpoint,
we might as well start with the lowest levels in the computational
hierarchy and work upward. In this paper we begin in just this way
by defining quantum versions of the simplest language classes --- the
regular and context-free languages \cite{hopcroft}.

To do this, we define quantum finite-state and push-down automata
(QFAs and QPDAs) as special cases of a more general object, a
{\em real-time quantum automaton}. In this setting a formal language
becomes a function that assigns quantum probabilities to words.
We also define {\em quantum grammars}, in which we sum over all
derivations to find the amplitude of a word.  We show that
the corresponding languages, generated by quantum grammars and
recognized by quantum automata, have pleasing properties in analogy to
their classical counterparts.  These properties include pumping lemmas,
closure properties, rational and (almost) algebraic generating
functions, and Greibach normal form.

For the most part, our proofs simply consist of tracking standard results
in the theory of classical languages and automata, stochastic automata, and 
formal power series, and attaching complex amplitudes to the transitions and 
productions of our automata and grammars.  In a few places --- notably, lemmas
12 and 13 and theorems 6, 7, 10, 19, 23, and 24 --- we introduce genuinely 
new ideas. 

We believe that this strategy of starting at the lowest rungs of the 
Chomsky hierarchy has several benefits.  First, we can make concrete 
comparisons between classical and quantum computational models. 
This comparison is difficult to make for more powerful models, 
because of unsolved problems in computer science (for instance, 
deterministic vs. quantum polynomial time).

Second, studying the computational power of a physical system can give
detailed insights into a natural system's structure and dynamics.
For example, it may be the case that the spatial density of physical 
computation is finite. In this case, every finite quantum computer
is actually a QFA.  If a system does in fact have infinite memory,
it makes sense to ask what kinds of long-time correlations it can have,
such as whether its memory is stack-like or queue-like. Our QPDAs
provide a way to formalize these questions.

Molecular biology suggests another example along these lines, the class
of protein secondary structures coded for by RNA. To some approximation
the long-range correlations between RNA nucleotide base pairs responsible
for secondary structure can be modeled by parenthesis-matching grammars
\cite{Searls,SakakibaraEtAl}. Since RNA macromolecules are quantum
mechanical objects, constructed by processes that respect atomic
and molecular quantum physics, the class of secondary structures coded
for by RNA may be more appropriately modeled by the quantum analogs
of context-free grammars introduced here. In the same vein, DNA and RNA
nucleotide sequences are recognized and manipulated by various active
molecules (e.g. transcription factors and polymerases), could their
functioning be modeled by QFAs and QPDAs?

Finally, the theory of context-free languages has been extremely useful
in designing compilers, parsing algorithms, and programming languages for 
classical computers.  Is it possible that quantum context-free languages
can play a similar role in the design of quantum computers and algorithms?

\subsection{Quantum mechanics}

First, we give a brief introduction to quantum mechanics \cite{qm}.

A quantum system's state is described by a vector of complex numbers.  
The dimension of a quantum system is the number of complex numbers
in its state vector.  A column vector is written $|a\ket$ and its
{\em Hermitian conjugate} $|a\ket^\dagger$, the complex conjugate
of its transpose, is the row vector $\bra a|$.  These vectors live
in a {\em Hilbert space} $H$, which is equipped with an inner product
$a \cdot b=\bra a|b \ket$.  The probability of observing a given state
$a$ is its norm $|a|^2=\bra a|a \ket$.

Over time, the dynamics of a quantum system rotates the state $|a\ket$
in complex vector space by a {\em unitary} matrix $U$ --- one whose inverse
is equal to its Hermitian conjugate, $U^\dagger=U^{-1}$.  Then the
total probability of the system is conserved, since if $\bra a'|=\bra a|U$, 
then $\bra a'|a' \ket = \bra a|U^\dagger U|a \ket = \bra a|a \ket$.

The eigenvalues of a unitary matrix are of the form $e^{i\omega}$, where 
$\omega$ is a real-valued angle, and so are restricted to the unit circle 
in the complex plane.  Thus, the dynamics of an $n$-dimensional 
quantum system, which is governed by an $n \times n$ unitary matrix, 
is simply a rotation in $\C^n$.  In the Schr\"odinger equation,
$U$ is determined by the {\em Hamiltonian} or energy operator $\cal H$
via $U=e^{i{\cal H}t}$.

A measurement consists of applying an operator $O$ to a quantum state $a$.
We will write operators on the right, $\bra a|O$.  To correspond to a 
classical observable, $O$ must be {\em Hermitian}, $O^\dagger=O$, so that 
its eigenvalues are real and so ``measurable''.
If one of its eigenvalues $\lambda$ is associated 
with a single eigenvector $u_\lambda$, then we observe the outcome 
$O=\lambda$ with a probability $|\bra a|u_\lambda \ket|^2$, where 
$\bra a|u_\lambda \ket$ is the component of $a$ along $u_\lambda$. 

More generally, if there is more than one eigenvector $u_\lambda$ with 
the same eigenvalue $\lambda$, then the probability of observing $O=\lambda$ 
when the system is in state $a$ is $|a P_\lambda|^2$, where $P_\lambda$ is a 
{\em projection} operator such that $\bra u_\mu|P_\lambda=\bra u_\mu|$ if 
$\mu=\lambda$ and $0$ otherwise.  Thus, $P_\lambda$ projects $a$ onto the 
subspace of $H$ spanned by the $u_\lambda$.

For instance, suppose that we consider a two-dimensional quantum system
with Hamiltonian ${\cal H}=\mat 1 & 0 \\ 0 & -1 \rix$.  Then
$U=\mat e^{it} & 0 \\ 0 & e^{-it} \rix$.  The eigenvectors of ${\cal H}$
are $\cho 1 \\ 0 \ose$ and $\cho 0 \\ 1 \ose$, with eigenvalues $+1$ and $-1$,
respectively.  If the system is in the state $\bra a|=(\sqrt{3}/2, -i/2)$, 
a measurement of the energy ${\cal H}$ will yield $+1$ or $-1$ with 
probabilities $3/4$ and $1/4$, respectively.  The projection operators are 
$P_{+1}=\mat 1 & 0 \\ 0 & 0 \rix$ and $P_{-1}=\mat 0 & 0 \\ 0 & 1 \rix$.

\subsection{Classical finite automata and regular languages}

Readers familiar with basic automata theory should skip this section
and the next two.  An introduction can be found in \cite{hopcroft}.

If $A$ is an {\em alphabet} or set of symbols, $A^*$ is the set of all
finite sequences or {\em words} over $A$ and a {\em language} $L$ over $A$
is a subset of $A^*$.  If $w$ is a word, then $|w|$ is its length and 
$w_i$ is its $i$'th symbol.  We denote the empty word by $\eps$, the 
concatenation of two words $u$ and $v$ as $uv$, and $w$ repeated 
$k$ times as $w^k$.

A {\em deterministic finite-state automaton} (DFA) consists of a 
finite set of states $S$, an input alphabet $A$, a transition function 
$F:S \times A \to S$, an initial state $s_\init \in S$, and a set of 
accepting states $S_\accept \subset S$.  The machine starts in $s_\init$
and reads an input word $w$ from left to right.  At the $i$th step, it
reads a symbol $w_i$ and updates its state to $s'=F(s,w_i)$. It accepts $w$
if the final state reached after reading $w_{|w|}$ is in $S_\accept$.
We say the machine {\em recognizes} the language of accepted words.

A {\em nondeterministic finite-state automaton} (NFA) has a transition 
function into the power set of $A$, $F:S \times A \to {\cal P}(A)$, 
so that there may be several transitions the machine can make for each 
symbol.  An NFA accepts if there is an allowed {\em computation path}, i.e.\
a series of allowed transitions, that leads to a state in $S_\accept$.

As it turns out, DFAs and NFAs recognize exactly the same languages,
since an NFA with a set of states $S$ can be simulated by a DFA whose 
states correspond to subsets of $S$.  If a language can be recognized by a
DFA or NFA, it is called {\em regular}.

For instance, the set of words over $A=\{a,b\}$ where no two $b$'s occur
consecutively is regular.  If $S=\{A,B,R\}$, $s_\init=A$, $S_\accept=\{A,B\}$,
and
\[ \begin{array}{cc} 
   F(A,a)=F(B,a)=A & F(A,b)=B \\
   F(B,b)=R & F(R,a)=F(R,b)=R 
\end{array} \]
then we enter the `reject' state $R$, and stay there, whenever we encounter
the string $bb$. $A$, $S$, $s_\init$, $S_\accept$, and $F$ constitute a DFA.

One way to view finite-state automata is with matrices and vectors.
If an NFA has $n$ states, the set of allowed transitions can be described 
by an $n \times n$ transition matrix $M_a$ for each symbol $a \in A$,
in which $(M_a)_{ij}=1$ if and only if the transition from state $i$ 
to state $j$ is allowed on reading $a$.  Then if $\vec{s}_\init$ is the 
$n$-component column vector 
\[ (\vec{s}_\init)_i = \left\{ \begin{array}{ll}
       1 & \quad i = s_\init \\
       0 & \quad \mbox{otherwise} \end{array} \right. \]
and $\vec{P}_\accept$ is the column vector
\[ (\vec{P}_\accept)_i = \left\{ \begin{array}{ll}
       1 & \quad i \in S_\accept \\
       0 & \quad \mbox{otherwise} \end{array} \right. \]
then the number of accepting paths on an input $w$ is
\beq f(w) = \vec{s}_\init^{\,T} \cdot M_w \cdot \vec{P}_\accept \eeq
where $M_w$ is shorthand for $M_{w_1} M_{w_2}\cdots M_{w_{|w|}}$.
Then a word $w$ is accepted if $f(w) > 0$, so that there is some path
leading from $\vec{s}_\init$ to the accepting subspace spanned by
$s \in S_\accept$.  (We apply the matrices on the right, so that they occur 
in the same order as the symbols of $w$, instead of in reverse.)  
Of course, $M_\eps$ is the identity matrix, which we will denote $\id$.

Equation (1) will be our starting point for defining quantum versions 
of finite-state automata and regular languages.

\subsection{Push-down automata and context-free languages}

A {\em push-down automaton} (PDA) is a finite-state automaton or `control' 
that also has access to a {\em stack}, an infinite memory storing
a string of symbols in some alphabet $T$.  Its transition function
$F:S \times T \times A \to {\cal P}(S \times T^*)$ allows it 
to examine its control state, the top stack symbol, and the input symbol.  
It then updates its control state, pops the top symbol off the stack, 
and pushes a (possibly empty) word onto the stack.  A PDA starts with 
an initial state and stack configuration. After reading a word, it
accepts if a computation path exists that either ends in an accepting
control state or produces an empty stack. 

PDAs recognize the {\em context-free} languages (CFLs), a name whose
motivation will become clear in a moment.  For instance, the Dyck language 
of properly nested words of brackets
$\{\eps, (), (()), ()(), (()()), \ldots\}$ is context-free.
It is recognized by a PDA with a single stack symbol $x$.  This PDA
pushes an $x$ onto the stack when it sees a ``$($'' and pops one
off when it sees a ``$)$''.  If it ever attempts to pop a symbol
off an empty stack, it enters the reject state and stays there.

A {\em deterministic push-down automaton} (DPDA) is one with at most one 
allowed transition for each combination of control state, stack symbol,
and input symbol. DPDAs recognize the {\em deterministic context-free}
languages (DCFLs), such as the Dyck language above.

\subsection{Grammars, context-free and regular}

A {\em grammar} consists of two alphabets $V$ and $T$, the {\em variables}
and {\em terminals}, an initial variable $I \in V$, and a set $P$ of 
{\em productions} $\alpha \to \beta$ where $\alpha \in V^*$ and 
$\beta \in (V \cup T)^*$.  A {\em derivation} $\alpha \To \beta$
is a chain of strings, where at each step one substring is replaced with 
another according to one of the productions.  Then the language generated by 
the grammar consists of those strings in $T^*$ (consisting only of terminals) 
that can be derived from $I$ with a chain of productions in $P$.

For example, the grammar $V=\{I\}$, $T=\{(,)\}$, and 
$P=\{I \to (I)I,\, I \to \eps \}$ generates the Dyck language.
Note that the left-hand side of each production consists of a single symbol 
and does not require any neighboring symbols to be present; hence the term 
{\em context-free}.  Context-free grammars generate exactly the languages 
recognized by PDAs.

The Dyck language grammar is {\em unambiguous} in that every word has a
unique derivation tree. A context-free language is unambiguous if there
is an unambiguous grammar that generates it. Notably, there are
{\em inherently ambiguous} context-free languages for which no unambiguous
grammar exists.

If we restrict a grammar further so that every production is of the form
$v_1 \to wv_2$ or $v_1 \to w$, where $w \in T^*$ and $v_1, v_2 \in V$, 
then there is never more than one variable present in the string.
The result is that a derivation leaves strings of terminals behind the
variable as it moves to the right. Such grammars are called {\em regular}
and generate exactly the regular languages.

\subsection{Quantum languages and automata}

Since quantum systems predict observables in a probabilistic way, 
it makes sense to define a {\em quantum language} as a function 
mapping words to probabilities, $f:A^* \to [0,1]$.  This generalizes the 
classical Boolean situation where each language has a {\em characteristic 
function} $\chi_L:A^* \to \{0,1\}$, defined as $\chi_L(w)=1$ if $w \in L$ 
and $0$ otherwise.  (In fact, in order to compare our quantum language classes 
with the classical ones, we will occasionally abuse our terminology by 
identifying a Boolean language with its characteristic function, saying 
that a language is in a given class if its characteristic function is.)

Then in analogy to equation (1), we define quantum automata in the 
following way:

\begin{definition}*  A {\em real-time quantum automaton} (QA) $Q$ consists of
\begin{itemize}
\item a Hilbert space $H$,
\item an initial state vector $s_\init \in H$ with $|s_\init|^2=1$,
\item a subspace $H_\accept \subset H$ and an operator $P_\accept$ 
that projects onto it,
\item an input alphabet $A$, and
\item a unitary transition matrix $U_a$ for each symbol $a \in A$.
\end{itemize}
Then using the shorthand 
\[ U_w = U_{w_1} U_{w_2} \cdots U_{w_{|w|}} , \]
we define the quantum language recognized by $Q$ as the function
\[ f_Q(w) = | s_\init U_w P_\accept |^2 \]
from words in $A^*$ to probabilities in $[0,1]$.  (Again, we apply
linear operators on the right, so that the symbols $w_i$ occur in 
left-to-right order.)
\end{definition}

In other words, we start with $\bra s_\init|$, apply the unitary matrices
$U_{w_i}$ for the symbols of $w$ in order, and measure the probability
that the resulting state is in $H_\accept$ by applying the projection
operator $P_\accept$ and measuring the norm.  This is a {\em real-time} 
automaton since it takes exactly one step per input symbol, with no additional 
computation time after the word is input.  

Physically, this can be interpreted as follows.  We have a quantum system 
prepared in a superposition of initial states.  We expose it over time to 
different influences depending on the input symbols, one time-step per symbol.  
At the end of this process, we perform a measurement on the system and $f(w)$ 
is the probability of this measurement having an acceptable outcome, such as 
being in a given energy level.

Note that $f$ is not a measure on the space of words.  It is the 
probability of a particular measurement after a given input.

This basic setting is not new.  If we restrict ourselves to real rather than
complex values and replace unitarity of the transition matrices with
{\em stochasticity} in which the elements of each row of the $U_a$ sum
to 1, we get the {\em stochastic automata} of Rabin \cite{rabin};
see also the review in \cite{macarie}.  If we generalize the $U_a$
to nonlinear maps in $\R^n$, we get {\em real-time dynamical recognizers} 
\cite{dynrec}.  If we generalize the $U_a$ to nonlinear Bayes-optimal update
maps of the $n$-simplex, we get {\em $\eps$-machine} deterministic
representations of recurrent hidden Markov models \cite{Crut92c,Uppe97a}.

Note that the effect of the matrix product $U_w=U_{w_1} U_{w_2} \cdots$
is to sum over all possible paths that the machine can take.  Each path
has a {\em complex amplitude} equal to the product of the amplitudes of the
transitions at each step.  Each of $U_w$'s components, representing 
possible paths from an initial state $s_0$ to a final state $s_{|w|}$, 
is the sum of these.  That is,
\[ (U_w)_{s_0,s_{|w|}} = \sum_{s_1,s_2,\ldots,s_{|w|-1}} (U_{w_1})_{s_0,s_1} 
     (U_{w_2})_{s_1,s_2} \cdots (U_{w_{|w|}})_{s_{|w|-1},s_{|w|}} \]
over all possible choices of the intervening states $s_1,\ldots,s_{|w|-1}$.
The difference from the real-valued (stochastic) case is that
{\em destructive interference} can take place.  Two paths can have
opposite phases in the complex plane and cancel each other out, 
leaving a total probability less than the sum of the two, since
$|a+b|^2 \le |a|^2+|b|^2$.  

Note that paths ending in different perpendicular states in $H_\accept$ 
add noninterferingly, $|a|^2+|b|^2$, while paths ending in the same state 
add interferingly, $|a+b|^2$.  This will come up several times in discussion 
below.

In analogy with Turakainen's generalized stochastic automata \cite{tura}
where the transition matrices do not necessarily preserve probability,
we will sometimes find it useful to relax unitarity:

\begin{definition}*  A {\em generalized real-time quantum automaton} is 
one in which the matrices $U_a$ are not necessarily unitary and the
norm of the initial state $s_\init$ is not necessarily 1.
\end{definition}

We can then define different classes of quantum automata by restricting the 
Hilbert space $H$ and the transition matrices $U_a$ in various ways:
first to the finite-dimensional case and then to an infinite memory
in the form of a stack.

\section{Quantum finite-state automata and regular languages}

The quantum analog of a finite-state machine is a system with a 
finite-dimensional state space, so

\begin{definition}*  A {\em quantum finite-state automaton} (QFA) 
is a real-time quantum automaton where $H$, $s_\init$, and the $U_a$
all have a finite dimensionality $n$.  A {\em quantum regular language} 
(QRL) is a quantum language recognized by a QFA.
\end{definition}

In this section, we will try to reproduce as many 
results as possible on classical regular languages in the quantum case.

\subsection{Closure properties of QRLs}

First, we define two operations on quantum automata that
allow us to add and multiply quantum languages.  The result is that
the set of QRLs is closed under these operations, just as 
stochastic languages are \cite{macarie,paz}.  

\begin{definition}*  If $u$ and $v$ are vectors of dimension $m$ and $n$,
respectively, their {\em direct sum} $u \oplus v$ is the $(m+n)$-dimensional
vector $(u_1,\ldots,u_m,v_1,\ldots,v_n)$.  If $M$ and $N$ are matrices, then 
$M \oplus N=\left(\!\begin{array}{c|c} M & 0 \\ \hline 0 & N 
\end{array}\!\right)$.

Then if $Q$ and $R$ are quantum automata with the same input alphabet, 
and if $a$ and $b$ are complex numbers such that $|a|^2+|b|^2=1$,
the {\em weighted direct sum} $aQ \oplus bR$ has initial state
$s'_\init=a s_\init^Q \oplus b s_\init^R$, projection operator
$P'_\accept=P_\accept^Q \oplus P_\accept^R$, and transition matrices 
$U'_a=U_a^Q \oplus U_b^R$.
\end{definition}

\begin{lemma}  If $Q$ and $R$ are QFAs and if $|a|^2+|b|^2=1$, then 
$aQ \oplus bR$ is a QFA and $f_{aQ \oplus bR}=|a|^2 f_Q + |b|^2 f_R$.
Therefore, if $f_1, f_2, \ldots, f_k$ are QRLs, then $\sum_{i=0}^k c_i f_i$
is a QRL for any real constants $c_i>0$ such that $\sum_{i=0}^k c_i=1$.
\end{lemma}

\begin{proof} Clearly 
$|s'_\init|^2=|a s_\init^Q|^2+|b s_\init^R|^2=|a|^2+|b|^2 = 1$.
The direct sum of two subspaces is a subspace, the direct sum
of unitary matrices is unitary, and the direct sum of two 
finite-dimensional quantum automata is finite-dimensional, so 
$aQ \oplus bR$ is a QFA.

Furthermore, $U'_w=U_w^Q \oplus U_w^R$ and
\[ f_{aQ \oplus bR}(w) = | a s_\init^Q U_w^Q P_\accept^Q |^2
   + | b s_\init^R U_w^R P_\accept^R |^2 
   = |a|^2 f_Q(w)+|b|^2 f_R(w) \]
(Note that the phases of $a$ and $b$ don't matter, only their norms.)
By induction we can sum any $k$ QRLs in this way, as long as 
$\sum_{i=0}^k c_i=1$.  
\qed \end{proof}

\begin{definition}*  If $u$ and $v$ are vectors of dimension $m$ and $n$,
respectively, then their {\em tensor product} $u \otimes v$ is the
$mn$-dimensional vector
$w_{\bra i,j \ket}=u_i v_j$ where $\bra i,j \ket=n(i-1)+j$, say, is a
pairing function.  If $M$ and $N$ are $m \times m$ and $n \times n$ matrices,
$M \otimes N$ is the $mn \times mn$ matrix 
$O_{\bra i,k \ket,\bra j,l \ket}=M_{ij} N_{kl}$.  Then if $Q$ and $R$
are quantum automata with the same input alphabet, $Q \otimes R$ is defined
by taking the tensor products of their respective $s_\init$, $P_\accept$, 
and the $U_a$.
\end{definition}

\begin{lemma}  If $Q$ and $R$ are QFAs, then $Q \otimes R$ is a QFA
and $f_{Q \otimes R}=f_Q f_R$.  Therefore, the product of any number of
QRLs is a QRL.
\end{lemma}

\begin{proof}  It is easy to show that if $a$ and $c$ are $m$-dimensional 
vectors and $b$ and $d$ are $n$-dimensional vectors, then
$\bra a \otimes b|c \otimes d \ket = \bra a|c \ket \bra b|d \ket$.
Therefore, $|s'_\init|^2=|s_\init^Q|^2 \, |s_\init^R|^2=1$.
The tensor product of finite-dimensional unitary matrices is unitary and
finite-dimensional, so $Q \otimes R$ is a QFA.

Furthermore, $U'_w=U_w^Q \otimes U_w^R$ and
\[ f_{Q \otimes R}(w) = | s_\init^Q U_w^Q P_\accept^Q |^2 \cdot
   | s_\init^R U_w^R P_\accept^R |^2 = f_Q(w) f_R(w) \]
By induction we can multiply any number of QRLs in this way.  
\qed \end{proof}

\begin{lemma}  For any $c \in [0,1]$, the constant function
$f(w)=c$ is a QRL.
\end{lemma}

\begin{proof}  Just choose any $s_\init$ and $P_\accept$ such that 
$|s_\init P_\accept|^2=c$, and let $U_a=\id$ for all $a$.  
\qed \end{proof}

Since we can add and multiply QRLs, we have

\begin{corollary}*  {\em Let $f_i$ be QRLs and let $c_i$ be a set of constants
such that $\sum_{i=0}^k c_i \le 1$. Then any polynomial $\sum_j c_j g_j$,
where each $g_j$ is a product of a finite number of $f_i$'s, is a QRL.}
\end{corollary}

In a sense, closure under (weighted) addition and multiplication are
complex-valued analogs of {\sc or} and {\sc and}.  Classical regular languages
are closed under both these Boolean operations, as well as complementation:

\begin{lemma}  If $f$ is a QRL, then $\of=1-f$ is a QRL.
\end{lemma}

\begin{proof}  Let $H'_\accept$ be the subspace of $H$ perpendicular 
to $H_\accept$ and $P'_\accept$ the projection operator onto it.  
Since $P_\accept+P'_\accept=\id$, $P_\accept P'_\accept=0$, the $U_w$
are unitary, and $|s_\init|^2=1$, we have
\bea 1 & = & |s_\init U_w|^2=|s_\init U_w (P_\accept+P'_\accept)|^2 \\
& = & |s_\init U_w P_\accept|^2 + |s_\init U_w P'_\accept|^2 \\
& = & f(w)+\of(w) \eea
where $\of(w)=|s_\init U_w P'_\accept|^2$.
\qed \end{proof}

Another property of classical regular languages is closure under
inverse homomorphism \cite{hopcroft}:

\begin{definition}*  A {\em homomorphism} $h:A^* \to A^*$ is a function that 
replaces symbols with words.  For instance, if $h(a)=b$ and $h(b)=ab$,
then $h(bab)=abbab$.  If $f$ is a quantum language, then its 
{\em inverse image} under $h$ is the language $(f \circ h)(w)=f(h(w))$.
(This looks wrong, but it is in fact the proper form for the characteristic
function of the inverse image of a set.  Formally, the mapping from sets
to characteristic functions acts like a contravariant functor.)  
\end{definition}

\begin{lemma}  If $f$ is a QRL and $h$ is a homomorphism,
then the inverse image $f \circ h$ is a QRL.
\end{lemma}

\begin{proof}  Simply replace each $U_a$ with $U_{h(a)}$.  Recall that
the composition of unitary matrices is unitary.
\qed \end{proof}

\subsection{The pumping lemma for QRLs}

The following is a well-known classical result \cite{hopcroft}:

\begin{lemma}*[(Pumping Lemma for Regular Languages)] If $L$ is a 
regular language, then any sufficiently long word $w \in L$ can be
written $w=xyz$ such that $xy^kz \in L$ for all $k \ge 0$.
\end{lemma}

\begin{proof}  If an NFA has $n$ states, then any path longer than $n$
transitions contains a loop, which can be repeated as many times 
as desired.  
\qed \end{proof}

Because of unitarity, we have a slightly stronger result for QRLs
in that any subword can be `pumped'.  However, unlike the classical case,
we can't repeat a word arbitrarily many times.  Rather, the dynamics
is like an irrational rotation of a circle, so that for any $\eps > 0$,
there is some $k$ such that $k$ rotations brings us back to within 
a distance $\eps$ from where we started.

\begin{theorem}[(Pumping for QRLs)]  If $f$ is a QRL, then for 
any word $w$ and any $\eps > 0$, there is a $k$ such that 
$|f(uw^kv)-f(uv)| < \eps$ for any words $u, v$.  Moreover, if 
$f$'s automaton is $n$-dimensional, there is a constant $c$ 
such that $k < (c\eps)^{-n}$.
\end{theorem}

\begin{proof}  In its diagonal basis, $U_w$ rotates $n$ complex numbers 
on the unit circle by $n$ different angles $\omega_i$ for $1 \le i \le n$.  
We can think of this as a rotation of a $n$-dimensional torus.  
If $V=(c\eps)^n$ is the volume of a $n$-dimensional ball of radius $\eps$, 
then $U_w^k$ is within a distance $\eps$ of the identity matrix for 
some number of iterations $k \le 1/V$.  We illustrate this in figure 1.

Then we can write $U_w^k=\id+\eps J$, where $J$ is a diagonal
matrix for which $\sum_{i=0}^n |J_{ii}|^2 \le 1$, and
\[ f(uw^kv)=|s_\init U_u (\id+\eps J) U_v P_\accept|^2
           =f(uv)+\eps |s_\init U_u J U_v P_\accept|^2 \]
Since 
\[ |s_\init U_u J U_v P_\accept|^2 \le |s_\init|^2
\sum_{i=0}^n |J_{ii}|^2  \le 1 \]
the theorem is proved.  
\qed \end{proof}

If $m$ of the angles $\omega_i$ are rational fractions $2\pi p/q$,
then we return to a $(n-m)$-dimensional torus every $q$ steps
and $k < q(c\eps)^{-(n-m)}$.

\begin{figure}
\centerline{\psfig{file=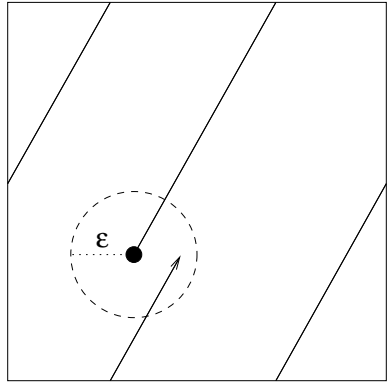,width=2in}}
\caption{Iterating the unitary matrix $U_w$ is equivalent to rotating
a torus.  If a ball of radius $\eps$ has volume $V$, then after at most
$1/V$ iterations the state must return to within a distance $\eps$ of
its initial position.}
\end{figure}

In the case where a unitary QFA recognizes a classical language 
(which we identify with its characteristic function), this gives
the following:

\begin{theorem}  If a regular language $L$ is a QRL, then the 
transition matrices $M_a$ of the minimal DFA recognizing $L$ 
generate a group $\{M_w\}$.  Therefore, there are regular languages 
that are not QRLs.
\end{theorem}

\begin{proof} Any set of matrices forms a semigroup, so we just have to 
show that every sequence of transitions $M_w$ has an inverse.  

Define two words as equivalent, $u \sim v$, if they can be followed 
by the same suffixes, $uw \in L$ if and only if $vw \in L$.  It is 
well-known \cite{hopcroft} that the states of $L$'s minimal DFA are in 
one-to-one correspondence with $\sim$'s equivalence classes.

Then if $L$'s characteristic function $\chi_L$ is a QRL, setting $\eps < 1$
in theorem 6 shows that for every $w$, there exists a $k$ such that,
for all $u$ and $v$,
\[ \chi_L(uw^kv) = \chi_L(uv) \]
which implies $uw^k \sim u$ for all $u$.  Then $M_w^k=\id$ in $L$'s minimal 
DFA since it returns any $u$ to its original equivalence class, and
$M_w$ has an inverse $M_w^{k-1}$.  So $\{M_w\}$ is a group.

Most regular languages don't have this property.  Consider the language 
$L$ given in the introduction with the subword $bb$ forbidden.  
Inserting $bb$ anywhere in an allowed word makes it disallowed, 
and this cannot be undone by following $bb$ with any other subword.
Thus $M_{bb}$ has no inverse in $\{M_w\}$, and $L$ is not a QRL.
\qed \end{proof}

In contrast, in the generalized case where the $U_a$ don't have to be
unitary, we have

\begin{lemma}  Any regular language is a generalized QRL.
\end{lemma}

\begin{proof} Let the $U_a$ be the Boolean transition matrices of $L$'s
DFA.  Then there is exactly one allowed path for each allowed word, so
$f(w)=\chi_L(w)$.
\qed \end{proof}

Combining this with the previous corollary gives the following:

\begin{corollary}*  The QRLs are a proper subclass of the generalized QRLs.
\end{corollary}

\subsection{QRLs are rational}

In classical language theory, we are often interested in the
{\em generating function} of a language, $g_L(z)=\sum_{w \in L} z^{|w|}$
or equivalently $\sum_n N_n z^n$, where $N_n$ is the number of words 
of length $n$ in $L$.  More generally, if we think of the symbols 
$a \in A$ as noncommuting variables, we can write a formal power series 
$G_L=\sum_{w \in L} w$, whereupon setting $a=z$ for all $a \in A$ gives 
$G_L=g_L(z)$.

A beautiful theory of such series is given in \cite{salomaa}.
In particular, the generating function of a regular language is always
{\em rational}, i.e.\ the quotient of two polynomials.  To see this,
sum equation (1) over all lengths, labelling transitions with their 
respective symbols.  Using a DFA with one computation path per word,
if we define $M=\sum_{a \in A} aM_a$ and rewrite the sum over all words 
as a sum over all lengths, we have 
\bea G_L & = & \sum_w \vec{s}_\init^{\,T} \cdot M_w \cdot \vec{P}_\accept \\
& = & \vec{s}_\init^{\,T} \cdot \sum_{n=0}^\infty M^n \cdot \vec{P}_\accept \\
& = & \vec{s}_\init^{\,T} \cdot (\id-M)^{-1} \cdot \vec{P}_\accept \eea
which is rational in each symbol $a$ since each component of $(\id-M)^{-1}$ is.
Then restricting to $a=z$ for all $a$ gives a rational $g_L(z)$ as well.

For instance, for the regular language given above with $bb$ forbidden,
$M=\mat a & b \\ a & 0 \rix$, $\vec{s}_\init=\cho 1 \\ 0 \ose$, and 
$\vec{P}_\accept=\id$.  Here $\cho 0 \\ 0 \ose$ represents the reject state.
Then the reader can check that
\[ (\id-M)^{-1}=\frac{1}{1-a-ab}\mat 1 & b \\ a & 1-a \rix \]
and
\[ G_L = \frac{1+b}{1-a-ab} = 1+a+b+aa+ab+ba+\cdots \]
where the empty word is now denoted by 1.  Setting $a=b=z$ gives
\[ g_L(z) = \frac{1+z}{1-z-z^2} = 1+2z+3z^2+5z^3+\cdots \]
recovering the well-known fact that the number of words of length $n$
is the $n$'th Fibonacci number.

The obvious generalization of this is

\begin{definition}*  If $f$ is a quantum language, then its 
{\em generating function} $G_f$ is the formal sum $\sum_{w \in A^*} f(w) \, w$.
\end{definition}

\begin{theorem}  If $f$ is a generalized QRL, then $G_f$ is rational.
\end{theorem}

\begin{proof}  We first consider generating functions $g$ based on 
complex amplitudes rather than total probabilities.  The accepting subspace 
$H_\accept$ is spanned by a finite number of perpendicular unit vectors $h_i$.  
Then if we define $g_i=\sum_w \bra s_\init|U_w|h_i\ket\, w$ and 
$U=\sum_{a \in A} aU_a$, we have
\[ g_i=\bra s_\init\,|\,(\id-U)^{-1}\,|\,h_i \ket \]
and the $g_i$ are rational.

The {\em Hadamard product} of two series $C=\sum_w c_w w$ and 
$D=\sum_w d_w w$ is the series formed by multiplying their coefficients 
term-by-term, $C \odot D=\sum_w c_w d_w w$.  Since 
$|v P_\accept|^2=\sum_i |\bra v|h_i\ket|^2$ for any vector $v$,
i.e.\ the probability of being in $H_\accept$ is the (noninterfering)
sum of the squares of the amplitudes along each of the $h_i$, we have
\[ G_f=\sum_i g_i^* \odot g_i \]
The class of rational series is closed under both addition and 
Hadamard product \cite{salomaa}, so $G_f$ is rational.  (These closure
properties are generalizations of the closure of the class of regular
languages under union and intersection.)
\qed \end{proof}

The theory of rational generating functions has also been used in the
recognition of languages by neural networks \cite{sontag}.

\subsection{Real representation and stochastic automata}

We should investigate the relationship between quantum and real-valued
stochastic automata, since the latter have been extensively studied.
We alluded to the following in the introduction \cite{paz,tura}:

\begin{definition}*  A {\em generalized stochastic function} is a 
function from words over an alphabet $A$ to real numbers, $f:A^* \to \R$, 
for which there are real-valued vectors $\pi$ and $\eta$ and real-valued
matrices $M_a$ for each $a \in A$ such that $f$ is a bilinear form,
\[ f(w)=\pi^T \cdot M_w \cdot \eta \]
where $M_w=M_{w_1} M_{w_2} \cdots M_{w_{|w|}}$ as before.  We will call
such a function {\em $n$-dimensional} if $\pi$, $\eta$ and the $M_a$
are $n$-dimensional.

If the components of $\eta$ are 0 and 1 denoting nonaccepting and
accepting states and if $\pi$ and the rows of the $M_a$ have non-negative
entries that sum to 1 so that probability is preserved, then $f$ is a 
{\em stochastic function}.  If we allow negative entries but still require that
$\pi$ and the rows of the $M_a$ sum to 1, then $f$ is {\em pseudo-stochastic}.
\end{definition}

It is well known that complex numbers $c=a+bi$ can be represented by 
$2 \times 2$ real matrices $\vec{c}=\mat a & b \\ -b & a \rix$. 
The reader can check that multiplication is faithfully reproduced 
and that $\vec{c}^T \vec{c}=|c|^2 \id$.  In the same way, an $n \times n$ 
complex matrix can be simulated by a $2n \times 2n$ real-valued matrix.
Moreover, this matrix is unitary if the original matrix is.

Using this representation, we can show the following:

\begin{theorem}  Any generalized QRL recognized by an $n$-dimensional 
generalized QFA is a $2n^2$-dimensional generalized stochastic function.
\end{theorem}

\begin{proof}  First we transform our automaton so that the output $f(w)$
is a bilinear, rather than quadratic, function of the machine's state.
As before, let $h_i$ be a set of perpendicular unit vectors spanning 
$H_\accept$.  Then
\bea f(w) & = & \sum_{i=0}^n |\bra s_\init \,|\, U_w \,|\, h_i \ket|^2 \\
& = & \sum_{i=0}^n \bra s^*_\init \otimes s_\init 
    \,|\, U^*_w \otimes U_w \,|\, h^*_i \otimes h_i \ket \\
& = & \bra s^*_\init \otimes s_\init \,|\, U^*_w \otimes U_w \,|\, 
    \sum_{i=0}^n h^*_i \otimes h_i \ket \eea
This has the form $\pi^T \cdot M_w \cdot \eta$ with 
$\pi=s^*_\init \otimes s_\init$, $M_a=U^*_a \otimes U_a$ for all $a \in A$, 
and $\eta=\sum_i h^*_i \otimes h_i$.  Since these are the tensor products 
of $n$-dimensional objects, they have $n^2$ dimensions.  However, their
entries are still complex-valued.

Using the representation above, we transform $\pi^T$, $M_a$, and $\eta$
into $2 \times 2n^2$, $2n^2 \times 2n^2$, and $2n^2 \times 2$ real-valued
matrices $\opi^T$, $\oM_w$, and $\oeta$, respectively, and 
$\opi^T \cdot \oM_w \cdot \oeta=f(w) \mat 1 & 0 \\ 0 & 1 \rix$.
Letting $\pi$ and $\eta$ be the top row of $\opi$ and the left column 
of $\oeta$, respectively, gives the desired real-valued, bilinear form.
\qed \end{proof}

This expression of a QRL as a generalized stochastic function gives us
transition matrices that are unitary but neither stochastic nor 
pseudo-stochastic.  A logical question, then, is whether the class of QRLs 
is contained in the class of stochastic functions, or vice versa, and 
similarly for the pseudo-stochastic functions.  Since the only matrices 
that are both pseudo-stochastic and unitary are permutation matrices, 
it seems more likely that the QRLs are incomparable with both these classes.
In that case, their intersection would be the stochastic quantum regular
languages (SQuRLs) \cite{diablo}.

If a generalized stochastic function $f$ is the characteristic function 
of some language $L$, then $L$ can be defined as $L=\{w \,|\, f(w) > 0 \}$.  
Turakainen \cite{tura} showed that $f$ can be replaced with a stochastic
function, in which case $L$ is a {\em 0-stochastic language}.  Bukharaev
\cite{buk} has shown that any such language is regular, so we have a 
converse to lemma 8:

\begin{corollary}*  If the characteristic function of a language $L$
is a generalized QRL, then $L$ is regular.
\end{corollary}

\section{Quantum context-free languages}

\subsection{Quantum push-down automata (QPDAs)}

Next, we define quantum push-down automata and show that several
modifications to the definition result in equivalent machines.

\begin{definition}*  A {\em quantum push-down automaton} (QPDA) is a
real-time quantum automaton where $H$ is the tensor product of a
finite-dimensional space $Q$, which we will call the {\em control state},
and an infinite-dimensional {\em stack space} $\Sigma$, each basis vector 
of which corresponds to a finite word over a stack alphabet $T$.  
We also require that $s_\init$, which is now infinite-dimensional,
be a superposition of a finite number of different initial control
and stack states.

Because of the last-in, first-out structure of a stack,
only certain transitions can occur.  If $q_1, q_2 \in Q$ are control
states and $\sigma_1, \sigma_2 \in T^*$ are stack states, then the
transition amplitude $\bra (q_1,\sigma_1) | U_a | (q_2,\sigma_2) \ket$ 
can be nonzero only if $t\sigma_1=\sigma_2$, $\sigma_1=t\sigma_2$, or 
$\sigma_1=\sigma_2$ for some $t \in T$.  In other words, transitions 
can only push or pop single symbols on or off the stack or leave the 
stack unchanged.  Furthermore, transition amplitudes can depend on
the control state and the stack, but only on the top (leftmost) symbol 
of $\sigma_1$ and $\sigma_2$, or on whether or not the stack is empty.

Finally, for acceptance we demand that the QPDA end in both an accepting
control state and with an empty stack.  That is, 
$H_\accept=Q_\accept \otimes \{\eps\}$ for some subspace 
$Q_\accept \subset Q$.
\end{definition}

This definition differs in several ways from that of classical PDAs 
\cite{hopcroft}.  First of all, the amplitude of a popping transition
can depend both on the top stack symbol and the one below it, since the
one below it is the top symbol of the stack we're making a transition to.
We do this for the sake of unitarity and time-symmetry, since the amplitude
of a pushing transition depends on both the top symbol and the symbol pushed.  
Similarly, popping transition amplitudes can depend on whether the stack 
will be empty afterwards.

In the generalized case where the transition matrices are not constrained
to be unitary, we can easily get rid of this dependence:

\begin{lemma}  A generalized QPDA can be simulated by a generalized QPDA 
whose transition amplitudes do not depend on the second-topmost stack symbol.
\end{lemma}

\begin{proof} Simply expand the stack alphabet to $T'=T \cup T^2$.
Let each stack symbol also inform the QPDA of the symbol below it
or that it is the bottom symbol.  For instance, the stack $stu$ 
becomes $(s,t)\,(t,u)\,u$.
\qed \end{proof}

However, we believe lemma 11 holds only in the generalized case.  
While the machine's dynamic is still unitary on the subset of the stack space
that we will actually visit, we see no way to extend it to the entire
stack space, including nonsense stacks like $(s,t)\,(u,w)$, in a unitary, 
time-symmetric way.  

Again, for time-symmetry's sake, since we can only pop one symbol at a time, 
we only allow ourselves to push one symbol at a time.  We next show that 
allowing us to push words of arbitrary length adds no additional power, 
just as for classical PDAs, at least in the generalized case:

\begin{lemma}  A generalized QPDA that is allowed to push words of
arbitrary length on the stack can be simulated by a generalized QPDA
as defined above, for which every move pushes or pops one symbol or
leaves the stack unchanged.
\end{lemma}

\begin{proof}  In the classical case, we can do this simply by
adding extra control states that push the word on one symbol at a time
(lemma 10.1 of \cite{hopcroft}).  However, this allows several steps per 
input symbol and thus violates our real-time restriction, so we need a 
slightly more subtle construction.

Suppose the old QPDA pushes words $\gamma$ of length at most $k$.
Then we expand the stack alphabet to composite symbols 
$T'=T^k \times \{1,\ldots,k\}$, which we will denote $(\beta,m)$,
and expand the set of control states to $Q'=Q \times \{1,\ldots,k\}$,
which we will denote $(q,m_0)$.

We represent the old QPDA's stack as shown in figure 2.  If the stack of 
the new QPDA is $(\beta_1,m_1)(\beta_2,m_2)\cdots(\beta_s,m_s)$, then each
$\beta_i$ represents a chunk of the old QPDA's stack, starting with 
$\beta_i$'s $m_{i-1}$'th symbol.  Alternately, each $m_i$ is a pointer
telling us to skip to the $m_i$'th symbol of $\beta_{i+1}$.
The pointer $m_0$ to $\beta_1$ is stored in the control state.

\begin{figure}
\centerline{\psfig{file=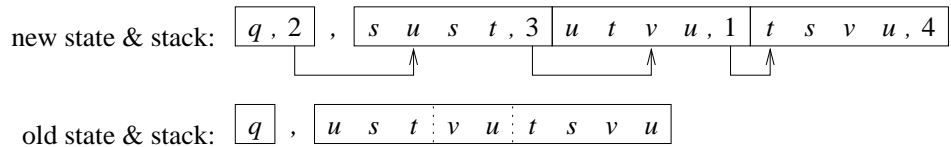,width=5in}}
\caption{Simulating a QPDA that can push words of length $\le 4$ on the stack
with one that only pushes or pops single symbols.  The counter $m_i$ in each
stack symbol $(\beta_i,m_i)$ acts as a pointer to the first relevant symbol 
in $\beta_{i+1}$. The pointer for $\beta_1$ is stored in the control state.
The symbols to the left of each pointer are either dummies or symbols that
have been popped off the original QPDA's stack.}
\end{figure}

Using lemma 11, we assume that the old QPDA's transition amplitudes
depend only on its top stack symbol.  We operate the new QPDA as follows, 
replacing the transitions of the old QPDA with new ones of the same amplitude:
\begin{itemize}
\item To pop the top symbol, i.e.\ the $m_0$'th symbol of $\beta_1$,
change the control state by incrementing $m_0$.  If $m_0=k$, pop 
$(\beta_1,m_1)$ off the stack and set $m_0=m_1$ in the control state.
\item To push a nonempty word $\gamma$ of length $n \le k$, choose 
a dummy symbol $a$ and push $(a^{k-n}\gamma,m_0)$ on the stack, padding 
$\gamma$ out to length $k$.  Then set $m_0=k-n+1$ in the control state.
\end{itemize}
This converts a QPDA into one where each transition pushes or pops one symbol,
or changes the topmost symbol of the stack by popping when $m_0=k$ and
then pushing a nonempty $\gamma$.  

This simulation preserves our real-time restriction, and creates a QPDA
which pushes or pops one symbol, or changes the top symbol, at each step.
To complete the proof, we need to convert this QPDA into one that
pushes, pops, or leaves the stack unchanged.  This can be done by making 
the top symbol part of the control state, $Q''=Q' \times T'$, so that 
we can change the top symbol by changing the state instead (as in 
lemma 10.2 of \cite{hopcroft}).  
\qed \end{proof}

Like lemma 11, we believe lemma 12 holds only in the generalized case.
Unitarity appears to be lost even on the set of stacks actually visited.
The stack state of the old QPDA is represented by many stack states
of the new QPDA, depending on the intervening computation, and 
some of these receive less probability than others.

In the classical case, acceptance by control state and by empty stack
are equivalent.  We can prove this in one direction, in both the
unitary and generalized case:

\begin{lemma}  If a quantum language is accepted by a (generalized) QPDA by 
empty stack, then it is accepted by a (generalized) QPDA by control state.
\end{lemma}

\begin{proof}  The standard construction (theorem 5.1 of \cite{hopcroft}) 
simply allows the PDA to empty its stack at the end of its computation, 
without reading any additional input.  Since this violates our real-time
restriction of one step per input symbol, we use a slightly different 
construction that also preserves unitarity.

First, double the number of control states to $Q'=Q \oplus \oQ$,
with a marked control state $\oq \in \oQ$ for each state $q \in Q$.  
Marked control states will denote an empty stack.
Then replace transitions of the old QPDA, that pop to or push on an 
empty stack, with new transitions, with the same amplitudes, as follows:
\begin{itemize}
\item Replace pops of the form $(q_1,t)\to(q_2,\eps)$ with 
$(q_1,t)\to(\oq_2,\eps)$
\item Replace pushes of the form $(q_1,\eps)\to(q_2,t)$ with 
$(\oq_1,\eps)\to(q_2,t)$
\end{itemize}
Require all states $(q,\eps)$ (an unmarked control state and an empty stack)
and $(\oq,\sigma)$ (a marked control state and a nonempty stack) to make 
transitions only to themselves with amplitude 1.  Finally, let $s_\init$
have nonzero components only along states $(\oq,\eps)$ that are marked and 
empty and $(q,\sigma)$ that are unmarked and nonempty.

Then the new QPDA will be in a marked control state if and only if the stack
is empty, so we accept with $H_\accept=\oQ_\accept \otimes \Sigma$.  
The new transition matrices are direct sums of the old ones 
(with the basis vectors $(q,\eps)$ replaced by $(\oq,\eps)$) with an 
identity matrix (on the space generated by the $(q,\eps)$ and 
$(\oq,\sigma)$).  Thus if the old QPDA is unitary, the new one is too.
\qed \end{proof}

Unfortunately, we believe that a QPDA accepting by control state 
without regard to the stack cannot in general be simulated by one
accepting by empty stack.  The accepting subspace 
$H_\accept=Q_\accept \otimes \Sigma$ is infinite-dimensional, 
allowing for an infinite number of different paths that add in a 
noninterfering way.  We see no way to map this into a finite-dimensional 
subspace of the form $Q_\accept \otimes \{\eps\}$.  Perhaps the reader can 
find a proof of this.

The last difference between QPDAs and classical PDAs is that, depending 
on its precise definition, a classical PDA either halts and accepts 
as soon as its stack becomes empty or rejects if it is asked to pop off an
empty stack.  In our case, we allow a QPDA to sense whether the stack is empty 
and act accordingly.  We do this because of our strict real-time constraint, 
in which the only time the QPDA is allowed to talk back to us is when we 
perform a measurement at the end of the input process.  Therefore, we have to 
tell the machine what to do if its stack is already empty and it receives 
more input.

\subsection{Quantum context-free grammars}

We now propose a definition of quantum grammars, in which each production 
has a set of complex amplitudes and multiple derivations of a word can 
interfere with each other constructively or destructively.
We show that in the context-free case, these grammars generate exactly the 
languages recognized by quantum PDAs.  

\begin{definition}*  A {\em quantum grammar} $G$ consists 
of two alphabets $V$ and $T$, the {\em variables} and {\em terminals}, 
an initial variable $I \in V$, and a finite set $P$ of {\em productions} 
$\alpha \to \beta$, where $\alpha \in V^*$ and $\beta \in (V \cup T)^*$.  
Each production in $P$ has a set of complex amplitudes $c_k(\alpha \to \beta)$ 
for $1 \le k \le n$, where $n$ is the {\em dimensionality} of the grammar.

We define the $k$'th amplitude $c_k$ of a derivation $\alpha \To \beta$ as
the product of the $c_k$'s for each productions in the chain and 
$c_k(\alpha \To \beta)$ as the sum of the $c_k$'s of all derivations
of $\beta$ from $\alpha$.  Then the amplitudes of a word $w \in T^*$ are
$c_k(w)=c_k(I \To w)$ and the probability associated with $w$ is the 
norm of its vector of amplitudes, summed over each dimension of the grammar,
$f(w)=\sum_{k=1}^n |c_k(w)|^2$.  We say $G$ {\em generates} the 
quantum language $f$.

Finally, a quantum grammar is {\em context-free} if only productions 
where $\alpha$ is a single variable $v$ have nonzero amplitudes. A 
{\em quantum context-free language} (QCFL) is one generated by some 
quantum context-free grammar.
\end{definition}

The main result of this section is that a quantum language is 
context-free if and only if it is recognized by a generalized QPDA.
We prove this with a series of lemmas that track the standard proof
almost exactly.  Our only innovation is attaching complex amplitudes to the
productions and transitions, and showing that they match.  A similar
proof in the real-valued case is given for {\em probabilistic tree automata}
in \cite{ellis}.

The multiple amplitudes $c_k$ attached to each production seem rather awkward.
As we will see below,
they are needed so that paths ending in perpendicular states in $Q_\accept$ 
can add in a noninterfering way.  If we had only one amplitude, then 
all paths would interfere with each other.  In the grammars we actually
construct, except for a few productions, the $c_k$'s for most will be equal.

\begin{definition}*  Two quantum grammars $G_1$ and $G_2$ are {\em equivalent} 
if they generate the same quantum language, $f_1(w)=f_2(w)$ for all $w$.
\end{definition}

\begin{definition}*  A quantum context-free grammar is in {\em Greibach
normal form} if only productions of the form $v \to a\gamma$ where 
$a \in T$ and $\gamma \in V^*$ can have nonzero amplitudes, i.e.\ 
every product $\beta$ consists of a terminal followed by a (possibly empty) 
string of variables.
\end{definition}

\begin{lemma}  Any quantum context-free grammar is equivalent to one in
Greibach normal form.
\end{lemma}

\begin{proof}  This is essentially the same proof as in \cite{ellis}
for the real-valued case.

Clearly $G'$ is equivalent to $G$ if for each derivation in $G$
of a terminal word, there is exactly one derivation in $G'$ with the same 
set of amplitudes. Then summing the amplitudes over all derivations will give
the same answer for both grammars.  All we need to do, then, is to attach
amplitudes to the standard proof for classical grammars (lemmas 4.1--4.4
and theorems 4.1--4.6 of \cite{hopcroft}) and show that they are carried
through correctly.  As shorthand, we will refer to $c_k$ and $c'_k$ 
for all $k$ as simply $c$ and $c'$, respectively.

First, theorem 4.4 of \cite{hopcroft} shows how to eliminate 
{\em unit productions} of one variable by another, $v_1 \to v_2$.  
If $G$ has such productions, then for every production 
$v_i \to \beta$ in $G$ where $\beta$ is not a single variable, 
give $G'$ the productions
\[ c'(v_i \to \beta) = c(v_i \To \beta)
   = \sum_j c(v_i \To v_j) \, c(v_j \to \beta) \]
for all $i$, where 
\[ c(v_i \To v_j)=\sum_{n=0}^\infty (M^n)_{ij}=(\id-M)^{-1}_{ij} \]
sums over all paths from $v_i$ to $v_j$ with $n$ unit productions,
and $M_{ij}=c(v_i \to v_j)$.  Then setting $c'(v_i \to v_j)=0$ 
leaves $G'$ with no unit productions.

Second, theorem 4.5 of \cite{hopcroft} converts a grammar
to {\em Chomsky normal form}, in which $\beta$ consists of either a 
single terminal or two variables.  For any production $v \to \beta$ in $G$
where $\beta$ consists of $m$ variables $b_1 b_2 \cdots b_m$, introduce
additional variables $d_1, d_2, \ldots d_{m-2}$ and allow the productions
$v \to b_1 d_1$, $d_1 \to b_2 d_2$, \ldots, $d_{m-2} \to b_{m-1} b_m$ 
in $G'$.  Then give $G'$ the productions
\[ c'(v \To \beta)=c'(v \to b_1 d_1)
   \cdot \prod_{i=1}^{m-3} c'(d_i \to b_{i+1} d_{i+1})
   \cdot c'(d_{m-2} \to b_{m-1} b_m) \]
which we can make equal to $c(v \to \beta)$ by choosing the $c'$ on the
right-hand site appropriately, e.g.\ with $c'(v \to b_1 d_1)=c(v \to \beta)$ 
and the others set to 1.

Finally, lemma 4.4 of \cite{hopcroft} eliminates productions of the form 
$v \to v \alpha$.  If $G$ has such productions and $v$'s other productions 
in $G$ are $v \to \beta$, add a variable $b$ and give $G'$ the productions
\bea
c'(b \to \alpha) = c'(b \to \alpha b) & = & c(v \to v \alpha) \\
c'(v \to \beta) = c'(v \to \beta b) & = & c(v \to \beta) \eea
for all $\alpha$ and $\beta$.  Then
\bea c'(v \To \beta \alpha_1 \alpha_2 \cdots \alpha_m)
   & = & c'(v \to \beta b) \cdot \prod_{i=1}^{m-1} c'(b \to \alpha_i b)
           \cdot c'(b \to \alpha_m) \\
   & = & c(v \to \beta) \cdot \prod_{i=1}^m c(v \to v \alpha_i) \\
   & = & c(v \To \beta \alpha_1 \alpha_2 \cdots \alpha_m) \eea
where the derivation tree for $G'$ now produces the $\alpha_i$ from 
left to right rather than from right to left.

The reader can easily check that the rest of the proof of theorem 4.6
of \cite{hopcroft} can be rewritten this way, so that $G$ and $G'$
have derivations with all the same complex amplitudes.
\qed \end{proof}

Greibach normal form is useful because the derivation trees it generates
create a terminal symbol on the left with every production.
Each such tree corresponds to a computation of a real-time PDA that
accepts with an empty stack.  Adding complex amplitudes
gives us the quantum version of theorem 5.3 of \cite{hopcroft}:

\begin{theorem}  Any QCFL is recognized by a generalized QPDA.
\end{theorem}

\begin{proof}  Convert the QCFL's grammar into Greibach normal form.
Then construct a QPDA with the terminals $T$ as its input symbols, with
the variables $V$ as its stack alphabet, and with one control state $q_k$ 
for each dimension of the grammar, $1 \le k \le n$.

Let the QPDA's transitions be as follows.  For each production 
$v \to a\gamma$ where $a \in T$ and $\gamma \in V^*$, 
if the control state is $q_k$ and the top stack symbol is $v$, 
let $U_a$ pop $v$ and push $\gamma$ on the stack with amplitude 
$c_k(v \to a\gamma)$.  Always leave the control state unchanged.

Then as we read the input symbols $a$, the QPDA guesses a derivation tree
and ends with an empty stack.  The amplitude of a computation path with 
control state $q_k$ is equal to the $k$'th amplitude of the corresponding 
derivation. Summing over all paths is equivalent to summing over all 
derivations.  If the QPDA's initial control state vector is 
$q_\init=(1,1,\ldots,1)$, the initial stack is $I$, and $Q_\accept=Q$, 
then projecting onto $H_\accept=Q \otimes \{\eps\}$ sums over all $k$ and 
gives the norm $f(w)=\sum_k |c_k(w)|^2$.

This gives us a QPDA that pushes whole words on the stack.  Using lemma 12, 
we can convert it into one that pushes or pops one symbol or leaves the 
stack unchanged, and we're done.
\qed \end{proof}

Conversely, by assigning the correct amplitudes to the productions 
in theorem 5.4 of \cite{hopcroft}, we can make each derivation match
a computation path of a QPDA:

\begin{theorem}  Any quantum language recognized by a generalized QPDA
is a QCFL.
\end{theorem}

\begin{proof}  By lemma 11, we will assume that the QPDA's transition 
amplitudes do not depend on the second-topmost stack symbol.

Our variables will be of the form $[q_1,t,q_2]$, where $q_1, q_2 \in Q$
and $t \in \Sigma \cup \{\eps\}$.  The leftmost variable will tell us that 
the QPDA is in control state $q_1$ with top symbol $t$ (or an empty stack
if $t=\eps$) 
and will be in state $q_2$ by the time $t$ is popped.  As in the previous 
theorem, the terminals will be the input symbols of the QPDA, and the
$k$'th amplitude $c_k$ of the derivation will be the amplitude of all paths
that end with a final state $q_k$.  Thus the dimensionality of the grammar
is equal to that of $Q_\accept$.

To start us off, we guess the QPDA's final state $q_k$, initial state $q_1$,
and initial stack $\beta$, and what states $q_2,\ldots,q_{|\beta|}$ 
we will go through as we pop the symbols of $\beta$.  For each allowed 
control state $q_k \in Q_\accept$, for each state-stack pair 
$(q_1,\beta)$ with nonzero amplitude in $s_\init$, and for all possible 
chains of control states $q_2,\ldots,q_{|\beta|} \in Q$, allow the production
\[ I \to [q_1,\beta_1,q_2]\,[q_2,\beta_2,q_3]
            \cdots[q_{|\beta|},\beta_{|\beta|},q_k] \]
with amplitudes $c_k=\bra s_\init|(q_1,\beta) \ket$ and $c_j=0$ 
for all $j \ne k$.  (These will be our only productions for which
$c_k$ depends on $k$.)

Then reading an input symbol $a \in A$, pushing a symbol $s$ on the stack, 
and entering state $q_3$ is represented by a production of the form 
\beq [q_1,t,q_2] \to a \, [q_3,s,q_4] \, [q_4,t,q_2] \eeq
whose amplitudes $c_k$ are all equal to the amplitude 
$\bra (q_1,\sigma)|U_a|(q_3,s\sigma) \ket$ of this QPDA transition.  
This production is allowed for any $q_4$, which is the state we guess 
that we will pass through after popping $s$ at some later time.

Similarly, reading an input symbol $a$, popping $t$ off the stack, and 
entering state $q_2$ is represented by
\beq [q_1,t,q_2] \to a \eeq
whose amplitudes $c_k$ are all equal to the amplitude
$\bra (q_1,t\sigma)|U_a|(q_2,\sigma) \ket$ of this transition.
Changing the state to $q_3$ while leaving the stack unchanged is 
represented by
\beq [q_1,t,q_2] \to a \, [q_3,t,q_2] \eeq
with amplitudes $c_k = \bra (q_1,\sigma)|U_a|(q_3,\sigma) \ket$.

Then, if we apply our productions always to the leftmost variable,
we see that each derivation tree corresponds to a computation path
of the QPDA with the same amplitude as the derivation.  Summing over 
derivations sums over computation paths.
$c_k(w) = \bra s_\init|U_a|(q_k,\eps) \ket$ is the amplitude
of all paths that end with the 
QPDA in control state $q_k$ with an empty stack.  Then 
$f(w)=\sum_{k=1}^n |c_k(w)|^2$ sums over all $q_k \in Q_\accept$
and the theorem is proved.
\qed \end{proof}

This representation of the control state, in which every control state
occurs in two variables, is necessary to enforce a consistent series of 
transitions, since symbols in a context-free derivation have no way of 
communicating with each other once they are created.

An alternate approach would be to give our productions {\em matrix-valued} 
amplitudes, so that their transitions can keep track of the state.
Our current definition, in which the $c_k$ are simply multiplied
componentwise, is equivalent to using diagonal matrices.  Since matrices 
do not commute in general, we would have to choose an order in which to 
multiply the production amplitudes to define a derivation's amplitude.  
A leftmost depth-first search of a derivation in Greibach normal form would 
still correspond to a computation path of a QPDA.
However, our proof of Greibach normal form breaks down because of the way 
lemma 4.4 of \cite{hopcroft} changes the shape of the tree.  If such grammars
can be put in Greibach normal form, then theorem 15 works and they are 
equivalent to QPDAs. If they cannot, they may be more powerful.

The productions in the above proof look nonunitary because they produce
either too much probability, since (2) is allowed for any choice of $q_4$,
or too little, since (3) and (4) may not correspond to transitions that are 
allowed at all.  Let us define

\begin{definition}*  A QCFL is {\em unitary} if it is recognized by a 
unitary QPDA.
\end{definition}

It is not clear what constraints a quantum grammar needs to meet to be unitary.
Nor is it clear whether these constraints can be put in a simple form that is 
preserved by the kinds of transformations we use in lemma 14.  Perhaps
a grammar's productions affect unitarity in a similar way to the
rule table of a quantum cellular automaton.  An algorithm to tell whether
a quantum CA is unitary is given in \cite{durr}.

Finally, we note that theorems 15 and 16 have the following corollaries:

\begin{corollary}*  Any quantum context-free grammar is equivalent to one 
in which the production amplitudes $c_k$ do not depend on $k$ except 
for productions from the initial variable.  Any generalized QPDA 
can be simulated by one whose transitions never change its control state,
for which $Q_\accept=Q$, and whose only initial stack consists of a 
single symbol.
\end{corollary}

It is not clear whether the latter is true in the unitary case.

\subsection{Closure properties of QCFLs}

Classical context-free languages are closed under intersection with a 
regular language.  The quantum version of this follows easily:

\begin{lemma}  If $f$ is a (unitary) QCFL and $g$ is a QRL, then $fg$ is a 
(unitary) QCFL.
\end{lemma}

\begin{proof}  We simply form the tensor product of the two automata.
If $f$ and $g$ have finite-dimensional state spaces $Q$ and $R$, 
construct a new QPDA with control states $Q \otimes R$, transition matrices 
$U'_a=U_a^f \otimes U_a^g$ (recall that $\otimes$ preserves unitarity), and 
accepting subspace $H'_\accept=Q_\accept \otimes R_\accept \otimes \{\eps\}$.
\qed \end{proof}

Classical CFLs are also closed under union, which as before becomes addition:

\begin{lemma}  If $f$ and $g$ are QCFLs, then $f+g$ is a QCFL.
\end{lemma}

\begin{proof}  We define a direct sum of two grammars as follows.
Suppose the grammars generating $f$ and $g$ have $m$ and $n$ dimensions,
variables $V$ and $W$, and initial variables $I$ and $J$.
We will denote their amplitudes by $c_k^f$ and $c_k^g$.
Then create a new grammar with $m+n$ dimensions, variables
$V \cup W \cup \{K\}$, and initial variable $K$, with the productions
$K \to I$ and $K \to J$ allowed with amplitudes $c_k=1$. Other
productions are allowed with $c_k=c_k^f$ for $1 \le k \le m$
and $c_k=c_{k-m}^g$ for $m+1 \le k \le m+n$.  The reader can easily check
that this grammar generates $f+g$.
\qed \end{proof}

We would like to say that a weighted sum $af+bg$, where $a+b=1$, 
of unitary QCFLs is unitary.  This is true if the QPDAs accepting
$f$ and $g$ have stack alphabets of the same size. Just take the 
direct sum of their control state spaces and let both sets of
states interpret the stack as if it were their own.  However, if one
stack alphabet is bigger than the other, we have to figure out how to
handle the dynamics in a unitary way when one of $f$'s states tries to 
read one of $g$'s stack symbols.  We leave this as a question
for the reader.

\subsection{The generating functions of QCFLs}

If we define a generating function of a context-free language $L$ that
counts multiple derivations, $G_L=\sum_{w \in L} n(w)\,w$, where $n(w)$ 
is the number of derivations of $w$ in $L$'s grammar, then $G_L$ is algebraic.
That is, it is a solution to a finite set of polynomial equations in 
noncommuting variables \cite{salomaa}.  If we don't count multiple 
derivations and define $G_L=\sum_{w \in L} w$ instead, then $G_L$ is 
algebraic for unambiguous context-free languages since each word has a 
unique derivation \cite{hopcroft}.

For instance, the Dyck language is generated by the unambiguous grammar
$P=\{I \to aIbI$, $I \to \eps\}$, where we have replaced left and right
brackets with $a$ and $b$ respectively.  Then its generating function
obeys the quadratic equation in noncommuting variables
\[ G = aGbG + 1 \]
If we set $a=b=z$, this becomes
\[ g(z) = z^2 g^2 + 1 \]
whose solution is
\[ g(z) = \frac{1-\sqrt{1-4z^2}}{2z^2} = 1+z^2+2z^4+5z^6+14z^8+\cdots \]
whose $z^{2k}$ coefficient is the {\em Catalan number} 
$\cho 2k \\ k \ose/\,(k+1)$.

The closest we can come to this in the quantum case is the following.

\begin{definition}*  The {\em Hadamard square} of a formal power series $g$
is the Hadamard product $g^* \odot g$.
\end{definition}

\begin{theorem}  If $f$ is a QCFL, then $G_f$ is a restriction of
the Hadamard square of an algebraic power series.
\end{theorem}

\begin{proof}  As in theorem 9, we start with generating functions weighted
with complex amplitudes rather than probabilities.  For each dimension $k$
of the grammar write $c$ for $c_k$ and define
\[ g_v = \sum_{w \in T^*} c(v \To w) \, w \]
This is the generating function of the terminal words $w \in T^*$ that 
can be derived from a variable $v \in V$, weighted by the $k$'th amplitudes
of each derivation.  For a terminal $a \in T$, we define $g_a=a$ since 
$a$ can only produce itself.  We also use the shorthand
\[ g_\beta = g_{\beta_1} g_{\beta_2} \cdots 
      g_{\beta_{|\beta|}} \]
since the words that can be derived from a word $\beta$ are simply
concatenations of those that can be derived from each of $\beta$'s symbols.

Then the $g_v$ obey the following equations, with one term for 
each production:
\[ g_v=\sum_{\beta \in (V \cup T)^*} c(v \to \beta) \, g_\beta \]
each of which is a polynomial of order 
$\max_{\beta|c(v \to \beta) \ne 0} |\beta|$.
This system of equations has an algebraic solution $g_I$.  

If we call the $g_I$ based on the $k$'th amplitude $g_k$, then $G_f$ is the 
sum of their Hadamard squares
\[ G_f = \sum_w f(w) \, w = \sum_w \sum_{k=1}^n |c_k(w)|^2 \, w 
        = \sum_{k=1}^n g^*_k \odot g_k \]
We can write this as a single Hadamard square in the following way.
For each dimension $k$ of the grammar, introduce a new symbol $x_k$.
Then if we define $g=\sum_{k=1}^n x_k g_k$, we have
\[ g^* \odot g = \sum_{k=1}^n x_k \, (g^*_k \odot g_k) \]
and $G_f=g^* \odot g$ in the restriction $x_k=1$ for all $k$.
\qed \end{proof}

Unfortunately, unlike the class of rational series, the class of algebraic 
series is not closed under Hadamard product.  This corresponds to the fact 
that the context-free languages are not closed under intersection.
In fact, the set of accepting computations of a Turing machine
is the intersection of two CFLs, so it is undecidable whether 
two algebraic series have a nonzero Hadamard product \cite{hopcroft}.

This also means that the Hadamard square of an algebraic series
can be transcendental.  Let $A$ and $B$ be two algebraic series 
such that $A \odot B$ is transcendental.  Then if $C=(A+B)/2$ and 
$D=(A-B)/2$, we have $A \odot B=(C \odot C)-(D \odot D)$ and 
at least one of $C \odot C$ and $D \odot D$ must be transcendental.  
As a concrete example, $g(z)=\sum_{z=0}^\infty \cho 2n \\ n \ose z^n$
is algebraic, but $g \odot g$ can be shown to be transcendental 
using the asymptotic techniques in \cite{flajolet}.

Ideally, this result could be used to show that certain inherently ambiguous
context-free languages, whose generating functions aren't the Hadamard square 
of an algebraic function, are not QCFLs.  Unfortunately, it is not obvious 
how to prove this, even in the case where all the $f(w)$ are 0 or 1.

\subsection{Regular grammars}

Although it is painfully obvious at this point, we include the following
for completeness.

\begin{definition}*  A quantum grammar is {\em regular} if only productions
of the form $v_1 \to wv_2$ and $v_1 \to w$ have nonzero amplitudes, where
$v_1, v_2 \in V$ are variables and $w \in T^*$ is a (possibly empty)
word of terminals.
\end{definition}

\begin{theorem}  A quantum language is a generalized QRL if and only if 
it is generated by a regular quantum grammar.
\end{theorem}

\begin{proof}  First we show that the language $f$ generated by a regular
quantum grammar is a generalized QRL.  Using the techniques of lemma 14, 
we can convert any regular grammar into one where $|w|=1$, i.e.\ 
all productions are of the form $v_1 \to av_2$ or $v_1 \to a$, where 
$v_1, v_2 \in V$ and $a \in T$.

If there are $m$ variables, then for each dimension $k$ of the grammar 
we can define a set of $(m+1)$-dimensional transition matrices $U^{(k)}_a$:
\[ (U^{(k)}_a)_{ij}=\left\{ \begin{array}{ll}
      c_k(v_i \to av_j) & \quad 1 \le i,j \le m \\
      c_k(v_i \to a)    & \quad j=m+1 \\
      0                 & \quad i=m+1
   \end{array} \right. \]
Then $|c_k(w)|=|s_\init U^{(k)}_w P_\accept|$, where $s_\init$ is the 
unit vector $(s_\init)_i=1$ if $v_i=I$ and 0 otherwise; and
$uP_\accept=u_{m+1}$, i.e.\ $P_\accept$ projects onto a vector's 
$(m+1)$'st component.  Then each $f_k=|c_k(w)|^2$ is a QRL and by lemma 1 
so is their sum $f(w)=\sum_{k=1}^n f_k(w)=\sum_{k=1}^n |c_k(w)|^2$.

Conversely, let $f$ be a generalized QRL.  Its state space is spanned 
by a set of unit vectors that we identify with the variables $V$.  
The accepting subspace $H_\accept$ is spanned by a set of unit vectors
$h_k$ as in theorem 9, each of which corresponds to one dimension
of the grammar.  Then define the production amplitudes as follows:
\bea 
c_k(I \to v)      & = & \bra s_\init|v \ket \\
c_k(v_i \to av_j) & = & (U_a)_{ij} \\
c_k(v_j \to \eps) & = & \bra v_j|h_k \ket
\eea 
Then $\sum_{k=1}^n |c_k(w)|^2=\sum_{k=1}^n |\bra s_\init|U_w|h_k \ket|^2
=|\bra s_\init |U_w| P_\accept \ket |^2$ and the theorem is proved.
\qed \end{proof}

Since only the last of the amplitudes in theorem 20 depend on $k$, 
we can add the following corollary:

\begin{corollary}*  Any regular grammar is equivalent to one in which
the $c_k$ don't depend on $k$ except for productions of the form 
$v \to \eps$.
\end{corollary}

Just as the regular languages are a proper subclass of the context-free
languages, we can show that the QRLs are a proper subclass of the QCFLs, 
in both the unitary and non-unitary cases:

\begin{theorem}  The QRLs are a proper subclass of the unitary QCFLs,
and the generalized QRLs are a proper subclass of the QCFLs.
\end{theorem}

\begin{proof}  Containment is given in both cases by using the control state 
of a (unitary) QPDA to simulate a (unitary) QFA while leaving its stack alone.  
It is proper because the language $L_=$ of words in $\{a,b\}$ with an 
equal number of $a$'s and $b$'s is a unitary QCFL (or rather, its 
characteristic function is) but not a generalized QRL, as we will now show.

Consider a QPDA with two control states $A$ and $B$ and one stack symbol $x$.  
The stack will indicate how many excess $a$'s or $b$'s we have, with the 
control state indicating which dominates.  Then starting with an empty stack 
$s_\init=(A,\eps)$, we can recognize $L_=$ with the transition matrices
\[ U_a = \begin{array}{c|cccccccc}
   & (A,\eps) & (A,x) & (B,x) & (A,xx) & (B,xx) & (A,xxx) & (B,xxx) & \cdots 
                                                                      \\ \hline 
(A,\eps) &    & 1     &       &        &        &         &         & \\
(A,x)    &    &       &       & 1      &        &         &         & \\
(B,x)    & 1  &       &       &        &        &         &         & \\
(A,xx)   &    &       &       &        &        & 1       &         & \\
(B,xx)   &    &       & 1     &        &        &         &         & \\
(A,xxx)  &    &       &       &        &        &         &         & \ddots \\
(B,xxx)  &    &       &       &        & 1      &         &         & \\
\vdots   &    &       &       &        &        & \ddots  &         &
\end{array} \] 
(with all other entries zero and $(B,\eps)$ left unchanged and unused) 
and $U_b=U_a^\dagger=U_a^{-1}$.  Since both $U_a$ and $U_b$ are unitary, 
this is a QPDA and $L_=$ is a unitary QCFL.

On the other hand, $L_=$'s generating function 
\[ g(z) = \sum_{n=0}^\infty \cho 2n \\ n \ose z^{2n} = \frac{1}{\sqrt{1-4z^2}} 
\]
is algebraic but not rational, so $L_=$ is not a generalized QRL by theorem 9.
\qed \end{proof}

Since regular grammars are also context-free, theorem 20 is another proof
that the generalized QRLs are a subclass of the QCFLs.

\subsection{QCFLs and CFLs}

Finally, we will compare our quantum classes to their classical counterparts.
Lemma 7 states that any regular language is a generalized QRL.  Similarly,
we have (again conflating a language with its characteristic function):

\begin{lemma}  Any unambiguous context-free language is a QCFL.  
More specifically, for any unambiguous CFL $L$ there is a quantum grammar 
of dimensionality $1$ such that $c(w)=\chi_L(w)$.
\end{lemma}

\begin{proof}  Simply give allowed and disallowed productions amplitudes
$1$ and $0$, respectively.  Since $L$ is unambiguous, each allowed word has
exactly one derivation, so $c(w)=\chi_L(w)$.  Since 0 and 1 are their own
squares, we also have $f(w)=|c(w)|^2=\chi_L(w)$.
\qed \end{proof}

Using the quantum effect of destructive interference, we can get the
following nonclassical result, showing that quantum context-free grammars
and QPDAs are strictly more powerful than classical ones:

\begin{theorem}  If $L_1$ and $L_2$ are unambiguous context-free languages,
their symmetric difference 
$L_1 \bigtriangleup L_2=(L_1 \cup L_2)-(L_1 \cap L_2)$ 
is a QCFL.
\end{theorem}

\begin{proof}  If $L_1$ and $L_2$ are generated by grammars with initial
variables $I_1$ and $I_2$, then create a new initial variable $I$ and allow
the productions $I \to I_1$ and $I \to I_2$ with amplitudes $1$ and $-1$,
respectively. Then $f=|c^{(1)}(w)+c^{(2)}(w)|^2=1$ if $w$ is in $L_1$ or
$L_2$, but not both.
\qed \end{proof}

\begin{corollary}*  There are QCFLs that are not context-free.
\end{corollary}

\begin{proof}  Let $L_1=\{a^i b^i c^j\}$ and $L_2=\{a^i b^j b^j\}$,
both of which are unambiguous context-free.  Then 
\[ L_1 \bigtriangleup L_2 = \{a^i b^j c^k \,|\, i=j \mbox{ or } j=k,
\mbox{ but not both}\} \]
is a QCFL, but it can be shown to be noncontext-free using the 
pumping lemma for context-free languages \cite{hopcroft}.
\qed \end{proof}

We can use interference in another amusing way:

\begin{theorem}  If $L_1$, $L_2$, and $L_3$ are unambiguous context-free
languages, then $(L_1 \cup L_2 \cup L_3)-(L_1 \cap L_2 \cap L_3)$ is a QCFL.
\end{theorem}

\begin{proof}  Create a new initial variable $I$ and allow the productions 
$I \to I_1$, $I \to I_2$, and $I \to I_3$ with amplitudes $1$, $e^{2\pi i/3}$,
and $e^{4\pi i/3}$, respectively.  Since these are $120^\circ$ apart, 
$f=|c^{(1)}(w)+c^{(2)}(w)+c^{(3)}(w)|^2$ if $w$ is in one or two, but
not all three, of the three languages.
\qed \end{proof}

Unfortunately, there are no sets of four or more vectors with norm 1 
such that the sum of any subset of them has norm 1, so this is as far as
this argument goes.\footnote{We are indebted to Jan-Christoph Puchta, 
David Joyner, Benjamin Lotto, and Dan Asimov for providing proofs 
of this fact.}

The next logical questions are whether all languages whose characteristic 
functions are QCFLs are context-sensitive \cite{hopcroft} and whether 
theorem 19 can be used to show that some inherently ambiguous CFLs, with 
transcendental generating functions, are not QCFLs.

\section{Conclusion and directions for further work}

We have defined quantum versions of finite-state automata, push-down
automata, and context-free grammars, and shown that many classical
results carry over into the quantum case.  We leave the reader with a 
set of open questions, some of which have already been mentioned above:
\begin{enumerate}
\item{What happens when we remove the real-time restriction, allowing
the machine to choose when to read an input symbol?  This adds no power
to classical DFAs and PDAs \cite{hopcroft}.  Does it in the quantum case?}
\item{What about two-way automata, that can choose to move left or right
on the input?  This adds nothing to classical DFAs \cite{hopcroft} or 
real-valued stochastic finite-state automata \cite{kaneps}.  Does it make
QFAs more powerful?}
\item{Is there a natural quantum analog of rational transductions 
\cite{berstel}, under which QRLs and QCFLs are closed without losing unitarity?}
\item{Are QRLs incomparable with stochastic and pseudo-stochastic functions?}
\item{Is each QRL recognized by a unique QFA (up to isomorphism) with the
minimal number of dimensions?  It might be possible to determine the 
eigenvalues of $U_w$ for all $w$ by Fourier analysis of $f(uw^kv)$.  
We could then reconstruct the $U_a$, since any set of matrices is determined 
by their eigenvalues and those of their products \cite{giles}.}
\item{Can grammars with noncommuting matrix-valued amplitudes be defined
in a consistent way and put in Greibach normal form?}
\item{Is there a simple way of determining whether a quantum context-free
grammar generates a unitary QCFL?}
\item{Can a QPDA be simulated by one that never changes its control state,
and for which $Q_\accept=Q$, without losing unitarity?}
\item{Is a weighted sum of unitary QCFLs a unitary QCFL, even when their 
QPDAs have stack alphabets of different sizes?}
\item{Is there a quantum analog to the Dyck languages $D_k$ and to
Chomsky's theorem that every CFL is a homomorphic image of the intersection 
of $D_k$ with a regular language?}
\item{Are the QCFLs contained in the context-sensitive languages?}
\item{Are there CFLs that are not QCFLs?}
\item{Can we define quantum versions of other real-time recognizer classes,
such as queue automata \cite{ccrm}, counter automata \cite{hopcroft}, and
real-time Turing machines \cite{benioff,deutsch}?}
\item{Are languages recognized by real-time QTMs the product of two QCFLs, 
analogous to intersection in the classical case \cite{hopcroft}?}
\item{We can easily define quantum context-sensitive grammars.  Do they
correspond to a quantum version of linear-bounded Turing machines 
\cite{hopcroft}?}
\end{enumerate}
We hope that quantum grammars and automata will be fruitful areas of research
and that they will be useful to people studying quantum computation.

\subsection{Acknowledgments}  

We are grateful to Bruce Litow, Philippe Flajolet, and Christophe Reutenauer
for the proof that the Hadamard square of an algebraic series can be
transcendental; Bruce Reznick, Jan-Christoph Puchta, Alf van der Poorten, 
Timothy Chow and Robert Israel for advice on rational generating functions; 
John Baez for pointing out reference \cite{giles}; Ioan Macarie and
Eduardo Sontag for a reading of the manuscript; Christian ``Ducky'' 
Reidys for help on functors; and Umesh Vazirani for helpful discussions.   
This work was supported at UC Berkeley by ONR Grant N00014-96-1-0524 and 
at the Santa Fe Institute by ONR Grant N00014-95-1-0975.

\end{document}